\pdfoutput=1

\documentclass[twocolumn,aps,prl,showpacs,superscriptaddress]{revtex4}

\usepackage{graphicx}
\usepackage{amsmath,amssymb}
\usepackage[colorlinks=true,citecolor=blue,urlcolor=blue]{hyperref}

\def\endproof{\vrule height6pt width6pt depth0pt}

\begin{document}



\title{Graph-Theoretic Approach to Quantum Correlations}


\author{Ad\'an Cabello}
 \email{adan@us.es}
 \affiliation{Departamento de F\'{\i}sica Aplicada II, Universidad de Sevilla, E-41012 Sevilla, Spain}

\author{Simone Severini}
 \email{simoseve@gmail.com}
 \affiliation{Department of Computer Science, and Department of Physics and Astronomy, University College London, WC1E 6BT London, United Kingdom}

\author{Andreas Winter}
 \email{andreas.winter@uab.cat}
 \affiliation{ICREA and F\'{\i}sica Te\`{o}rica: Informaci\'{o} i Fenomens Qu\`{a}ntics, Universitat Aut\`{o}noma de Barcelona, E-08193 Bellaterra (Barcelona), Spain}


\date{\today}



\begin{abstract}
Correlations in Bell and noncontextuality inequalities can be expressed as a positive linear combination of probabilities of events. Exclusive events can be represented as adjacent vertices of a graph, so correlations can be associated to a subgraph. We show that the maximum value of the correlations for classical, quantum, and more general theories is the independence number, the Lov\'asz number, and the fractional packing number of this subgraph, respectively. We also show that, for any graph, there is always a correlation experiment such that the set of quantum probabilities is exactly the Gr\"otschel-Lov\'asz-Schrijver theta body. This identifies these combinatorial notions as fundamental physical objects and provides a method for singling out experiments with quantum correlations on demand.
\end{abstract}



\maketitle


{\em Introduction.---}Quantum theory (QT) is the basis of our current description of nature and is arguably the most successful theory in the history of science. However, we still do not understand QT in the sense that we cannot single out QT as the only theory that satisfies a set of principles similar to the two principles from which special relativity is derived \cite{Feynman65}.

In the search for similar principles for QT, much effort has been devoted to seeking principles singling out quantum nonlocal correlations (i.e., those that cannot be explained with local theories) \cite{PR94,V05,PPKSWZ09,NW09}. However, this emphasis on nonlocality assumes that experiments involving spacelike separated tests are more fundamental than other types of correlation experiments, while nothing in the rules of QT supports this assumption.

A more general approach to quantum correlations follows from the question of whether there exists a joint probability distribution that gives the marginals predicted by QT. This question is equivalent to the question of whether a specific set of linear correlation inequalities is satisfied \cite{Fine82a,Fine82b,Pitowsky89,KCBS08,AQBTC13}. For experiments with spacelike separated tests, these inequalities are called Bell inequalities \cite{Bell64,CHSH69} and for more general scenarios are called noncontextuality (NC) inequalities \cite{KCBS08,MWZ00,CFHR08,Cabello08,SBKTP09,BBCP09,KZGKGCBR09,ARBC09,LLSLRWZ11,YO12,KBLGC12,ZUZAWDSDK13,DHANBSC13,AACB13}. The experimental violation of NC inequalities reveals contextual correlations that cannot be explained with theories in which outcomes are predefined and do not depend on which combination of jointly measurable observables is considered (i.e., noncontextual theories \cite{Peres95,LKGC11}).

In this Letter we present a novel approach to quantum correlations within this more general framework of contextual correlations. We use graphs to characterize correlations and we show that three different classes of theories, namely, noncontextual theories, QT, and more general probabilistic theories, allow for different sets of probabilities. We will consider theories that assign probabilities to ``events'' defined as follows. We assume that preparations of physical systems are reproducible and can be compared through their statistics with respect to the available experiments. Preparations that yield the same probabilities for each of the experiments are considered equivalent and define the same {\em state}. Reciprocally, experiments that yield the same statistics for all states are considered equivalent and define the same {\em test}. Sets of tests and their corresponding outcomes that occur with the same probability in either state are considered equivalent and define the same {\em event}: outcome $a$ for test $x$ and outcomes $b,\ldots,c$ for tests $y,\ldots,z$ are equivalent if probabilities $P(a|x)$ and $P(b,\ldots,c|y,\ldots,z)$ are equal in either state. Then, $a|x$ and $b,\ldots,c|y,\ldots,z$ will denote two possible realizations of the same event. Two events $e_i$ and $e_j$ are {\em exclusive} if there exist two jointly measurable observables $\mu_i$, defined by $e_i$, and $\mu_j$, defined by $e_j$, that distinguish between them.


\begin{figure}[b]
\centering
\vspace{-9.0cm}
\centerline{\includegraphics[scale=0.45]{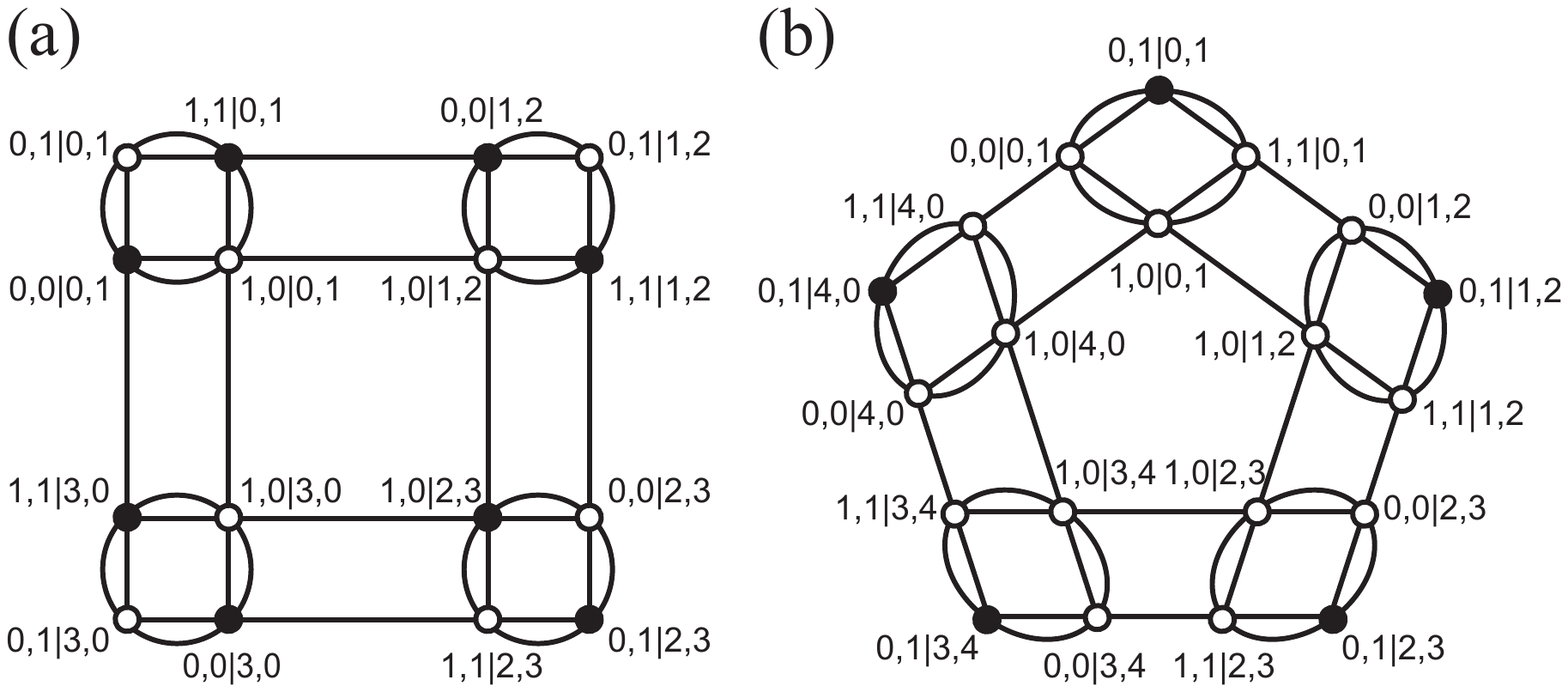}}
\vspace{-0.4cm}
\caption{\label{Fig1} (a) Simplified representation of the exclusivity graph of the CHSH experiment, ${\cal G}_{\rm CHSH}$. (b) {\em Idem} of the KCBS experiment, ${\cal G}_{\rm KCBS}$. Events are represented by vertices. Here, for simplicity, sets of pairwise exclusive events are represented by vertices in the same straight line or circumference rather than by cliques. (a) The exclusivity graph of $S_{\rm CHSH}$, denoted as $G_{\rm CHSH}$, is the induced subgraph of ${\cal G}_{\rm CHSH}$ obtained by removing all but the eight black vertices. An induced subgraph is obtained by selecting a subset of vertices and their incident edges. We use $G$ instead of $(G,w)$ whenever vertex weights are all~1. $G_{\rm CHSH}$ is isomorphic to the eight-vertex circulant $(1,4)$ graph $Ci_8(1,4)$. (b) The exclusivity graph of $S_{\rm KCBS}$, denoted as $G_{\rm KCBS}$, is the induced subgraph of ${\cal G}_{\rm KCBS}$ obtained by removing all but the five black vertices. $G_{\rm KCBS}$ is isomorphic to a five-cycle $C_5$ (i.e., a pentagon).}
\end{figure}


{\em Exclusivity graphs.---}To any correlation experiment, we can associate a graph ${\cal G}$ in which events are represented by vertices and pairs of exclusive events are represented by adjacent vertices. We will refer to ${\cal G}$ as the {\em exclusivity graph of the experiment}.

For example, in the Clauser-Horne-Shimony-Holt (CHSH) Bell experiment \cite{CHSH69} there are four tests $0,1,2,3$, each of them with two possible outcomes: $0$ and $1$. Tests $0$ and $2$ can be performed by Alice, and tests $1$ and $3$ by Bob. The experiment consists of performing the pairs of tests $(0,1)$, $(1,2)$, $(2,3)$, and $(3,0)$ on systems in the same quantum state. The exclusivity graph of the CHSH experiment ${\cal G}_{\rm CHSH}$ is represented in Fig.~\ref{Fig1} (a). It has 16~vertices and 12~cliques of size~4 (sets of four pairwise adjacent vertices).

Similarly, in the Klyachko-Can-Binicio\u{g}lu-Shumovsky (KCBS) contextuality experiment \cite{KCBS08,AACB13} there are five tests $i=0,\ldots,4$ with two possible outcomes $0$ and $1$, and the experiment consists of performing the five pairs of tests $(i,i+1)$, with the sum modulo~5, on systems in the same quantum state. The exclusivity graph of the KCBS experiment ${\cal G}_{\rm KCBS}$ is shown in Fig.~\ref{Fig1} (b). It has 20~vertices and 15~cliques of size~4.

The correlations in any Bell or NC inequality are expressed as a linear combination of probabilities of a subset of events of the corresponding experiment.
The fact that the sum of probabilities of outcomes of a test is~1 can be used to express these correlations as a {\em positive} linear combination of probabilities of events, $S=\sum_i w_i P(e_i)$, with $w_i>0$. For example, the CHSH and KCBS inequalities can be expressed \cite{BBCGL11}, respectively, as
\begin{subequations}
\begin{align}
 &S_{\rm CHSH}=\sum_{i=0}^3 \sum_{a,b} P(a,b\,|\,i,i+1) \stackrel{\mbox{\tiny{ LHV}}}{\leq} 3,
 \label{CHSH}\\
 &S_{\rm KCBS}=\sum_{i=0}^4 P(0,1\,|\,i,i+1) \stackrel{\mbox{\tiny{ NCHV}}}{\leq} 2,
 \label{KCBS}
\end{align}
\end{subequations}
where the second sum in Eq.~(\ref{CHSH}) is extended to $a,b \in \{0,1\}$ with $a=b$ if $i \neq 2$ and $a \neq b$ if $i=2$, the sum in $i+1$ is taken modulo~4 in Eq.~(\ref{CHSH}) and modulo~5 in Eq.~(\ref{KCBS}), and LHV and NCHV denote local and noncontextual hidden variables, respectively. Although in these examples all probabilities have weight 1, each probability $P(e_i)$ may have a different weight $w_i$. A vertex-weighted graph $(G,w)$ is a graph $G$ with vertex set $V$ and weight assignment $w:V \rightarrow \mathbb{R}_+$.

We can associate to $S$ a vertex-weighted graph $(G,w)$, where $G \subseteq {\cal G}$ and $i \in V$ represents event $e_i$ such that $P(e_i)$ is in $S$, adjacent vertices represent exclusive events, and the vertex weights represent the weights $w_i$ of the probabilities $P(e_i)$. We will call $(G,w)$ the {\em exclusivity graph of $S$}. The exclusivity graphs of $S_{\rm CHSH}$ and $S_{\rm KCBS}$ are represented in Figs.~\ref{Fig1} (a) and (b), respectively.


In order to define a general class of theories assigning probabilities to events, we will consider theories satisfying the following principle: The sum of probabilities of any set of pairwise exclusive events cannot be higher than~1. This class has been previously considered in \cite{Specker60,Wright78}. Specker noticed that classical and QT satisfy this principle, but that there are theories that do not \cite{Specker60,LSW11}. Following \cite{Cabello13,Yan13}, we will refer to this principle as the {\em exclusivity principle}. We will denote by $E1$ those theories satisfying the exclusivity principle applied to $G$ alone. The index~1 in $E1$ is used to distinguish these theories from those satisfying the exclusivity principle applied jointly to $G$ and other independent graphs \cite{Cabello13,Yan13}.

We first show that the exclusivity graph of $S$ can be used to calculate the limits of the correlations in classical, quantum, and theories satisfying $E1$.


{\em Result~1:} Given $S$ corresponding to a Bell or NC inequality, the maximum value of $S$ for classical (LHV and NCHV) theories, QT, and theories satisfying $E1$ is given by
\begin{equation}
 S \stackrel{\mbox{\tiny{ LHV,NCHV}}}{\leq} \alpha(G,w) \stackrel{\mbox{\tiny{{\em Q}}}}{\leq} \vartheta(G,w) \stackrel{\mbox{\tiny{ {\em E}1}}}{\leq} \alpha^*(G,w),
 \label{main}
\end{equation}
where $\alpha(G,w)$ is the independence number of $(G,w)$ \cite{GLS86}, $\vartheta(G,w)$ is the Lov\'asz number of $(G,w)$ \cite{Lovasz79,GLS81,GLS86}, and $\alpha^*(G,w)$ is the fractional packing number of $(G,w)$ \cite{GLS81,GLS86,Shannon56}. $\vartheta(G,w)$ might be only an upper bound to the maximum quantum value of $S$ in cases in which the particular physical settings of the experiment testing $S$ add further constraints.


{\em Proof:} The maximum value of $S$ for classical theories is always reached by a model in which each of the tests has a predefined outcome, i.e., a model in which each of the events in $S$ has either probability 0 or 1. Since mutually exclusive events cannot both have probability~1, the maximum value of $S$ for classical theories occurs when probability 1 is assigned to each vertex in a set of nonadjacent vertices in $G$. A set of nonadjacent vertices is called an independent or stable set of vertices of $G$.

Therefore, the maximum of $S$ for classical theories is the maximum of $\sum_i w_i$ where the maximum is taken over all stable sets of vertices of $G$. This is exactly the independence number of $(G,w)$, denoted as $\alpha(G,w)$ \cite{GLS86}.

An orthonormal representation (OR) in $\mathbb{R}^d$ of a graph $G$ with vertex set $V$ assigns a nonzero unit vector $|v_i \rangle \in \mathbb{R}^d$ to each $i \in V$ such that $\langle v_i | v_j \rangle=0$ for all pairs $i,j$ of nonadjacent vertices. A further unit vector $|\psi \rangle \in \mathbb{R}^d$, called {\em handle} \cite{Lovasz79}, is sometimes specified together with the OR. The definition of OR does not require that different vertices be mapped onto different vectors nor that adjacent vertices be mapped onto nonorthogonal vectors.
The complement $\overline{G}$ of a graph $G$ is the graph with vertex set $V$ such that two vertices $i,j$ are adjacent in $\overline{G}$ if and only if $i,j$ are not adjacent in $G$.
The Lov\'asz number of $(G,w)$ can be defined \cite{GLS86} as
\begin{equation}
 \vartheta(G,w):= \max \sum_{i\in V} w_i | \langle \psi | v_i \rangle |^2,
 \label{Lovasz}
\end{equation}
where the maximum is taken over all ORs of $\overline{G}$ and handles in any dimension.

Taken into account that the maximum value of $S$ in QT is always obtained for a quantum pure state $|\psi\rangle$ and a set of projectors $\Pi_i$ in a real Hilbert space of suitable dimension, the fact that $\vartheta(G,w)$ equals the maximum value of $S$ in QT is evident by noticing that $\langle \psi | \Pi_i | \psi\rangle$ can be written as $| \langle \psi | v_i \rangle |^2$, where $|v_i \rangle = \Pi_i |\psi\rangle/\sqrt{\langle \psi |\Pi_i| \psi \rangle}$ for all $i \in V$ is an OR of $\overline{G}$ and the handle $|\psi\rangle$ is the quantum state leading to the maximum value of $S$ in QT.

The maximum value of $S$ for theories satisfying $E1$ is proven by recalling that the fractional packing number (or fractional stability number) of $(G,w)$ \cite{GLS86,Shannon56} is defined as
\begin{equation}
 \alpha^*(G,w):=\max \sum_{i \in V} w_i p_i,
\end{equation}
where the maximum is taken over all $p_i \geq 0$ and for all cliques $C$ of $G$, under the restriction $\sum_{i \in C} p_i \leq 1$. This last restriction imposes that the sum of the probabilities of any set of pairwise exclusive events in $S$ cannot exceed~1.\hfill \endproof


{\em Comments:} Computing $\alpha(G,w)$ for arbitrary graphs is NP (nondeterministic polynomial time) complete even when all vertex weights are 1 \cite{GLS86,GLS88}, which is in agreement with the well-known result that computing the upper
bound for correlation inequalities for classical theories is NP-hard. The Lov\'asz number of $G$ was introduced \cite{Lovasz79} as an upper bound to the independence number and the Shannon capacity of $G$ \cite{Shannon56} (which is not even known to be computable). The extension of the Lov\'asz number to weighted graphs was introduced in \cite{GLS81,GLS86}. For any vertex weights, the Lov\'asz number can be computed to any desired precision in polynomial time \cite{GLS81,GLS88}. Computing $\alpha^*(G,w)$ for arbitrary graphs is NP-hard \cite{GLS81,GLS88} because we need the list of cliques of $G$. If these are given, then $\alpha^*(G,w)$ is a linear program, and as such efficiently computable.


{\em Examples:} When we apply Result~1 to $S_{\rm CHSH}$, we obtain $\alpha(G_{\rm CHSH})=3$, $\vartheta(G_{\rm CHSH})=2+\sqrt{2}$, and $\alpha^*(G_{\rm CHSH})=4$, which correspond to the maximum for local \cite{CHSH69}, quantum \cite{Tsirelson80}, and no-signaling theories \cite{PR94}, respectively.

When we apply Result~1 to $S_{\rm KCBS}$, we obtain $\alpha(G_{\rm KCBS})=2$, $\vartheta(G_{\rm KCBS})=\sqrt{5}$, and $\alpha^*(G_{\rm KCBS})=\frac{5}{2}$, which correspond to the maximum for noncontextual \cite{KCBS08}, quantum \cite{BBCGL11}, and no-disturbance theories \cite{RSKD12}, respectively.

The KCBS inequality can be extended to any odd number of settings $n \ge 5$, and then Result~1 leads to
\begin{equation}
 \sum_{i=0}^{n-1} P(0,1|i, i+1) \stackrel{\mbox{\tiny{ NCHV}}}{\leq} \frac{n-1}{2} \stackrel{\mbox{\tiny{{\em Q}}}}{\leq} \frac{n \cos(\pi/n)}{1+\cos(\pi/n)} \stackrel{\mbox{\tiny{ {\em E}1}}}{\leq} \frac{n}{2},
\end{equation}
where the sum in $i+1$ is taken modulo $n$. This inequality was also obtained in \cite{LSW11}.


For a given Bell or NC inequality (i.e., for a given $S$), $\vartheta(G,w)$ only provides an upper bound to its quantum maximum \cite{SBBC13}. Then, a natural question is whether, given $G$, there is a NC inequality that reaches $\vartheta(G)$.


{\em Result 2:} For any graph $G$, there is always a NC inequality such that the quantum maximum is {\em exactly} $\vartheta(G)$ and the set of quantum probabilities is {\em exactly} the Gr\"otschel-Lov\'asz-Schrijver theta body ${\rm TH}(G)$ \cite{GLS86}.


{\em Proof:} Given $G$, by Eq.~(\ref{Lovasz}), there is always an OR of $\overline{G}$ in $\mathbb{R}^d$, $\{|v_i\rangle\}$, and a handle $|\psi\rangle$ such that $\vartheta(G)= \max \sum_{i\in V} | \langle \psi | v_i \rangle |^2$. Let $D$ be the minimum dimension $d$ in which this OR exists. Then, consider the following positive linear combination of probabilities of events:
\begin{equation}
 S = \sum_{i\in V} P(1,0,\ldots,0|i,i_1,\ldots,i_{n(i)}),
\end{equation}
where test $i$ is defined as $\Pi_i=|v_i\rangle \langle v_i|$, with $|v_i\rangle$ in the OR, test $i_j$ is $\Pi_{i_j}=|v_{i_j}\rangle \langle v_{i_j}|$, with $|v_{i_j}\rangle$ in the OR, and $\{i_1,\ldots,i_{n(i)}\}$ is the set of tests corresponding to vertices adjacent to $i$. This $S$ reaches $\vartheta(G)$ when a quantum system is prepared in the quantum state $|\psi\rangle$. Notice that $1,0,\ldots,0|i,i_1,\ldots,i_{n(i)}$, which belongs to the same equivalence class as $1|i$, is a repeatable event with a well-defined probability, even though $i,i_1,\ldots,i_{n(i)}$ need not be jointly measurable.

In QT, the set of probabilities that can be assigned to the vertices of $G$ by performing tests on a quantum system is
\begin{widetext}
\begin{equation}
 {\cal Q}(G):= \left\{ (|\langle \psi | v_i \rangle |^2 : i \in V) : (|v_i\rangle : i \in V)\text{ is an OR of }\overline{G}\text{ and }|\psi\rangle \text{ a handle}\right\}.
\end{equation}
\end{widetext}
If the quantum system has dimension greater than or equal to $D$, then ${\cal Q}(G)$ is exactly the theta body ${\rm TH}(G)$ introduced in \cite{GLS86} [see Theorem 3.5 in \cite{GLS86} and Corollary 9.3.22 (c) in \cite{GLS88}].\hfill \endproof


{\em Comments:} Result~2 identifies $\vartheta(G)$ as a fundamental physical limit for quantum correlations and ${\rm TH}(G)$ as {\em the} set of {\em physical} correlations for a given $G$. Result~2 suggests that an important question for understanding quantum correlations is which is the principle that singles out ${\rm TH}(G)$ among all possible sets of probabilities that can be assigned to the vertices of $G$.

Although here we have shown that, for quantum physics, ${\rm TH}(G)$ is a fundamental set, ${\rm TH}(G)$ was originally introduced to bound the size of independent sets \cite{GLS86}.

Notice that Result~2 is not longer true if we replace ``NC inequality'' by ``Bell inequality.'' For example, if $G$ is a pentagon, there is no Bell inequality reaching $\vartheta(G)$ \cite{SBBC13}. This is due to the extra constraints imposed by the Bell scenario which enforce a specific labeling of the events (see \cite{SBBC13} for details). This shows the advantage of discussing quantum correlations in the framework of NC inequalities not referring to any specific experimental scenario.

Notice that this strategy of focusing on graphs without referring to any specific experimental scenario (whose existence is guaranteed by Result~2) substantially simplifies the problem of characterizing the quantum set with respect to the case in which the labeling is given \cite{Tsirelson80,Tsirelson93,NPA07,NPA08}.


In a similar way that we have defined the quantum set ${\cal Q}(G)$, we can define the corresponding sets for classical and more general theories satisfying $E1$. They are, respectively,
\begin{subequations}
\begin{align}
 &{\cal C}(G):=\text{convex hull} \left\{x^S : x^S \text{ is a stable labeling of }G\right\},\\
 &{\cal E}^1(G):= \left\{p \in \mathbb{R}^{|V|}_+ : \sum_{i\in C} p_i \leq 1\text{ for all cliques }C\right\},
\end{align}
\end{subequations}
where a stable labeling of $G$ is a labeling $x^S\in \{0,1\}^{|V|}$ for a stable set $S$ of $G$ such that $x^S_i =1$ if $i \in S$ and $x^S_i=0$ if $i \notin S$. Clearly, the quantum set is sandwiched between the classical and the $E1$ set,
\begin{equation}
{\cal C}(G) \subseteq {\cal Q}(G) \subseteq {\cal E}^1(G).
\end{equation}

A natural question is which graph properties distinguish quantum from classical correlations. Another interesting question is when quantum correlations are singled out by $E1$. Notice that ${\cal C}(G)$ and ${\cal E}^1(G)$ are polytopes, but ${\cal Q}(G)$, in general, is not.


{\em Result 3:} (i) ${\cal C}(G) = {\cal Q}(G)$ if and only if $G$ has no odd cycle $C_n$, with $n \ge 5$, or its complement $\overline{C_n}$ as induced subgraphs.

(ii) ${\cal Q}(G) = {\cal E}^1(G)$ if and only if $G$ has no odd cycle $C_n$, with $n \ge 5$, or its complement $\overline{C_n}$ as induced subgraphs.

(iii) ${\cal Q}(G)$ is a polytope if and only if $G$ has no odd cycle $C_n$, with $n \ge 5$, or its complement $\overline{C_n}$ as induced subgraphs.

{\em Proof:} ${\cal C}(G)$ is exactly the stable set polytope of $G$, ${\rm STAB}(G)$ (also called vertex packing polytope) \cite{GLS86,GLS88}. ${\cal E}^1(G)$ is exactly the fractional stable set polytope of $G$, ${\rm QSTAB}(G)$ (or clique-constrained stable set polytope) introduced in \cite{Shannon56}.

In \cite{GLS86} it is proven that ${\rm STAB}(G)={\rm TH}(G)$ if and only if $G$ is perfect. Perfect graphs were introduced in \cite{Berge61} in connection to the problem of the zero-error capacity of a graph \cite{Shannon56}. The strong perfect graph theorem \cite{CRST06} states that $G$ is perfect if and only if $G$ has no odd cycle $C_n$, with $n \ge 5$, or its complement $\overline{C_n}$ as induced subgraphs. This proves (i).

In \cite{GLS86} it is proven that ${\rm TH}(G)={\rm QSTAB}(G)$ if and only if $G$ is perfect. This proves (ii).

Finally, in \cite{GLS86} it is proven that ${\rm TH}(G)$ is a polytope if and only if $G$ is perfect \cite{GLS86}. This proves (iii).\hfill \endproof


{\em Applications.---}Results 1 and 2 provide a general method to construct NC inequalities. Specifically, they show that, for every graph $G$ such that $\alpha(G)<\vartheta(G)$, there is a NC inequality with classical limit given by $\alpha(G)$ and quantum violation given by $\vartheta(G)$ and that, for every $G$ such that $\alpha(G)<\vartheta(G)=\alpha^*(G)$, there is a NC inequality in which the maximum quantum violation cannot be higher without violating the exclusivity principle. This allows us to design experiments with quantum contextuality on demand by selecting graphs with the desired relationships between these three numbers.

The fact that these three numbers have been studied for a long time and that there is an extensive literature on the subject also opens the possibility of identifying novel interesting quantum correlations. For example, it is known that for arbitrarily large $n$ there are graphs for which $\alpha(G) \approx 2 \log n$ and $\vartheta(G) \approx \sqrt{n}$ or for which $\alpha(G)=3$ and $\vartheta(G) \approx \sqrt[4]{n}$ \cite{Peeters96}. This shows that the quantum violation of NC inequalities can be arbitrarily large. In addition, a question such as which are the correlations with maximum quantum contextuality can now be addressed, since now it can be related to the question of which graphs have the maximum $\vartheta(G)/\alpha(G)$ for a given number of vertices. Similarly, all possible forms of quantum contextuality can be classified by classifying graphs according to their combinatorial numbers.


{\em Conclusions.---}Here we have introduced a graph-based approach to the study of quantum correlations. First, we have shown that we can associate a graph ${\cal G}$ to any correlation experiment such that the possible correlations are given by the possible probability distributions that can be assigned to the vertices of the graph. Hence, the correlations considered in any Bell or NC inequality can be associated to a weighted subgraph $(G,w)$ such that $G \subseteq {\cal G}$. We have shown that the limits imposed to the correlations in classical, quantum, and more general theories can be obtained from three combinatorial numbers characteristic of $(G,w)$.

Then we have shown that, reciprocally, given any graph $G$, there is always a correlation experiment in which the whole set of quantum probabilities for $G$ can be reached. This result leads to identify the set of quantum probabilities for $G$ as a fundamental physical object and suggests that a fundamental question is to find the principle that singles out this set.

Our results provide a general method to construct NC inequalities, identify experimental scenarios with correlations on demand by picking out graphs with the required properties, and classify quantum correlations through the study of their graph properties.


{\em Note added.---}Preliminary versions of some of the results in this Letter were in a preprint \cite{CSW10} that has by now inspired numerous further developments.


\begin{acknowledgments}
 We thank B. Amaral, M. Ara\'ujo, C. Budroni, O.~G{\"u}hne, M. Kleinmann, J.-\AA. Larsson, A. J. L\'opez-Tarrida, J. R. Portillo, and M. Terra Cunha for useful conversations. A.C.'s work is supported by Project No.\ FIS2011-29400 with FEDER funds (MINECO, Spain), the FQXi large grant project ``The Nature of Information in Sequential Quantum Measurements,'' and the Brazilian program Science without Borders. S.S.'s work is supported by the Royal Society and EPSRC. A.W.'s work is supported by Project No.\ FIS2008-01236 with FEDER funds (MINECO, Spain), the European Commission (STREP ``QCS''), the European Research Council (Advanced Grant ``IRQUAT''), and the Philip Leverhulme Trust.
\end{acknowledgments}



\end{document}